\documentclass[aip,jcp,amsmath,amssymb,reprint]{revtex4-1}

\usepackage{graphicx}% Include figure files
\usepackage{dcolumn}% Align table columns on decimal point
\usepackage{bm}% bold math
\usepackage[utf8]{inputenc}
\usepackage[T1]{fontenc}
\usepackage{mathptmx}
\usepackage{tikz}
\usepackage{import}
\graphicspath{{plots/}{.}}
\usepackage{pgf}
\usepackage{pgfplots}
\usepackage{caption, subcaption}
\usepackage{floatflt}
\usepackage{todonotes}
\usepackage{hyperref}
\usepackage[capitalize]{cleveref}
\usepackage{siunitx}

\newcommand\inputpgf[2]{{
		\let\pgfimageWithoutPath\pgfimage
		\renewcommand{\pgfimage}[2][]{\pgfimageWithoutPath[##1]{#1/##2}}
		\input{#1/#2}
}}

\newcommand{\Pec}{\mathrm{Pe}}
\newcommand{\tum}{\mathrm{t}}
\newcommand{\run}{\mathrm{r}}

\newcommand{\dif}{\mathrm{d}}
\newcommand{\Dr}{D_\mathrm{r}}
\newcommand{\Dt}{D_\mathrm{t}}

\newcommand{\Tc}{T_\mathrm{c}}
\newcommand{\Tr}{T_\mathrm{r}}
\newcommand{\Tt}{T_\mathrm{t}}
\newcommand{\qr}{q_\mathrm{r}}
\newcommand{\qt}{q_\mathrm{t}}

\newcolumntype{Y}{>{\centering\arraybackslash}X}
\newcommand{\Us}{U_\mathrm{s}}
\newcommand{\Ua}{U_\mathrm{eff}}

\begin{document}

\preprint{AIP/123-QED}

\title[A Computational Model for Bacterial Run-and-Tumble Motion]{A Computational Model for Bacterial Run-and-Tumble Motion}

\author{Miru Lee}
\email{mlee@icp.uni-stuttgart.de}
\affiliation{Institute for Computational Physics, University of Stuttgart, Allmandring 3, 70569 Stuttgart, Germany}

\author{Kai Szuttor}
\affiliation{Institute for Computational Physics, University of Stuttgart, Allmandring 3, 70569 Stuttgart, Germany}

\author{Christian Holm}
\email{holm@icp.uni-stuttgart.de}
\affiliation{Institute for Computational Physics, University of Stuttgart, Allmandring 3, 70569 Stuttgart, Germany}

\date{\today}

\begin{abstract}
  In our article we present a computational model for the simulation
  of self-propelled anisotropic bacteria. To this
  end we use a self-propelled particle model and augment it with a
  statistical algorithm for the run-and-tumble motion.  We derive an equation for the
  distribution of reorientations of the bacteria that we use to
  analyze the statistics of the random walk and that allows
  us to tune the behavior of our model to the characteristics of an
  \textit{E.coli} bacterium. We validate our implementation in terms of a single
  swimmer and demonstrate that our model is capable of
  reproducing \textit{E. coli}'s run-and-tumble motion with excellent
  accuracy.
\end{abstract}

\maketitle

\section{Introduction}
%Due to its incredible abundance in Nature, \textit{E. coli} has become one of the mostly used bacteria in studying biological systems.
Living organisms like bacteria have developed several strategies to
enable their survival. One of the strategies that are particular to
flagellated bacteria is the so-called run-and-tumble (RT) motion,
which helps them to explore the surroundings and find food. Such
run-and-tumble bacteria, \textit{e.g.}, \textit{E. coli}, swim
straight for a certain amount of time and rather abruptly change the
swimming directions~\cite{berg72a,berg93a}. However, in models for
bacteria this behavior is often neglected or coarse-grained out by
using a stochastic description fixing only the diffusion coefficients,
which results in ignoring the precise information contained in the
spatial trajectories of RT bacteria. Another downside of approaches
such as active Brownian dynamics is that they do not include
hydrodynamical interactions. Although a stochastic description is one
of the most powerful tools to study and understand a bacterial
system~\cite{ebeling99a,elgeti13a,elgeti15a,ezhilan15b,cates13a,cates12a},
these downsides can miss important physical interactions when studying
the collective behavior of bacteria in complex environments.

There have been studies on individual bacteria, mainly on the
hydrodynamic interactions of a bacterium with its surroundings~\cite{lighthill52a,blake71b,winkler16a,lauga11a,zoettl16a}. Since
the flow fields induced by bacteria decay rather slowly
($\sim r^{-2}$)~\cite{lighthill52a,blake71b,winkler16a}, the inclusion
of hydrodynamical interactions seems to be a necessary ingredient in
modeling bacterial motion.  In this article we therefore present a
novel numerically efficient implementation of an elongated
self-propelled bacteria model that performs a RT motion and is able to
hydrodynamically interact with other bacteria and complex obstacles or
interfaces.

The article is organized as follows.  In~\cref{sec:theory}, we review
experimental discoveries and theoretical studies of RT
motion~\cite{berg72a,berg93a,saragosti12a,lovely75a}. Then, we derive
a formula that can be used to analyze the trajectory of a RT motion.
In~\cref{sec:imple} we introduce a molecular dynamics (MD) force-free
swimmer model that is implemented by coupling the bacterium to a
lattice-Boltzmann algorithm
(LB-MD)~\cite{duenweg09a,ahlrichs99a,krueger17a,succi01a,degraaf16a,degraaf16b}.
Further, we describe our method that is running on top of this swimmer
model that steers the RT motion. This will enable us to efficiently study
a system with multiple interacting run-and-tumble bacteria. The
algorithm itself is, however, not bounded to a certain simulation
method, so it can run on top of any conventional numerical
scheme. In~\cref{sec:results}, we present and analyze the trajectory
results for a single swimmer. There, we fix the relevant RT
algorithmic parameters of the swimmer to match the characteristics of
\textit{E. coli} bacteria and demonstrate that our model can reproduce
the experimentally observed swimming trajectories.

%Many studies of a bacterial system in a fluid have been done in two different
%ways. One way would be to focus on individual behaviors of a bacterium,
%studying mainly the hydrodynamic interactions of a bacterium with a
%fluid~\cite{lighthill52a,blake71b,winkler16a,lauga11a,zoettl16a}. The other
%way would be to focus on collective dynamics of bacteria. Such studies, more
%often than not, ignores the above-mentioned hydrodynamic
%interactions~\cite{ebeling99a,elgeti13a,elgeti15a,ezhilan15b}. However, the
%exclusion of hydrodynamics can be justified if and only if the bacterial
%density of a given system is so low that the bacteria rarely run into each
%other. Nevertheless, the flow fields induced by bacteria decay rather slowly
%($\sim r^{-2}$)~\cite{lighthill52a,blake71b,winkler16a}, so the inclusion of
%hydrodynamics when studying collective bacterial dynamics is worth
%considering.

\section{Statistical theory of the run-and-tumble motion}\label{sec:theory}
In the following, we build a mathematical model for bacterial RT
motion by means of statistical theory. A RT motion is characterized by the
following three distributions: the durations of runs, tumbles and
reorientations.

Throughout the statistical derivation we assume that
\begin{itemize}
	\item{the swimmer does not change its direction while running,}
	\item{the swimmer keeps the swimming speed constant while running,}
	\item{the swimmer does not move forward while tumbling.}
\end{itemize}
\subsection{Distributions}\label{subsec:dist}
It is well known that the durations of runs and tumbles follow a Poisson
statistics~\cite{berg72a,berg93a}. The probability mass function (Pm) for
a RT swimmer to terminate its current state of motion and transit to the other
motion (running $\leftrightarrow$ tumbling) for a given number of trials $k$ is provided by the so-called geometric
distribution function, which is nothing but a discrete version of the
exponential distribution
function~\cite{lovely75a,saragosti12a,berg72a,berg93a}:
\begin{equation}
\begin{aligned}
\mathrm{Pm}(k;q_{\mathrm{r/t}}) = (1-q_{\mathrm{r/t}})^{k-1}q_{\mathrm{r/t}},\qquad&\textrm{for }k = 1,2,3,\cdots,\\
&\textrm{with } 0<q_{\mathrm{r/t}}<1,
\end{aligned}
\label{eq:geometric_dist}
\end{equation}
where $q_{\mathrm{r/t}}$ is the termination rate for a given state: r for \textbf{r}uns and t for \textbf{t}umbles. For example, $\qt$ is the termination rate for tumbling, transitioning to the running phase.

The average number of trials $k_{\mathrm{r/t}}$ is
\begin{equation}
\left<k_{\mathrm{r/t}}\right>_k = \frac{1}{q_{\mathrm{r/t}}},
\label{eq:geo_1mo}
\end{equation}
where $\left<...\right>_k$ denotes an average over $k$.
The termination rate $q_{\mathrm{r/t}}$ can hence be
deduced from the average number of trials $\left<k_{\mathrm{r/t}}\right>_k$.

Poisson statistics demands an additional parameter that is
not present in~\cref{eq:geometric_dist}: the time step $\delta t$ between two
successive trials. In the following we will refer to this time step as the Poisson time step, which essentially defines the time resolution in measuring the durations of runs and tumbles. The detailed discussion can be found in~\cref{subsec:pram,sec:results}.

Experimentally obtained values for the durations are, therefore, given by the
following expression:
\begin{equation}
\left<T_{\mathrm{r/t}}\right>_k = \delta t \left<k_{\mathrm{r/t}}\right>_k = \frac{\delta t}{q_{\mathrm{r/t}}},
\label{eq:ave_time_resol}
\end{equation}
\textit{i.e.}, $\qr$ and $\qt$ correspond to the average duration of runs
$\left<T_\run\right>_k$ and that of tumbles $\left<T_\tum\right>_k$,
respectively.

The second moment $\left<k^2_{\mathrm{r/t}}\right>_k$ for the number of trials is given by:
\begin{equation}
\left<k^2_{\mathrm{r/t}}\right>_k =\frac{2-q_{\mathrm{r/t}}}{q_{\mathrm{r/t}}^2},
\label{eq:geo_2mo}
\end{equation}
and the standard deviation $\sigma_{k_{\mathrm{r/t}}}$ is thus given by
\begin{equation}
    \sigma_{k_{\mathrm{r/t}}} = \sqrt{\left<k_{\mathrm{r/t}}^2\right>_k-\left<k_{\mathrm{r/t}}\right>_k^2} = 
        \sqrt{\frac{1-q_{\mathrm{r/t}}}{q_{\mathrm{r/t}}^2}} \overset{q_{\mathrm{r/t}} \ll 1}{\approx} \frac{1}{q_{\mathrm{r/t}}} =
        \left<k_{\mathrm{r/t}}\right>_k.
\label{eq:geo_std}
\end{equation}
Note that if $q_{\mathrm{r/t}}\ll 1$, the standard deviation is approximated by the average number of trials.

The reorientation distribution can be described by
a random walk on the surface of a
sphere~\cite{berg72a,berg93a,saragosti12a}. Such a random walk can be formulated
via Fick's law in spherical coordinates~\cite{saragosti12a,berg93a}:
\begin{equation}
\partial_t p(\theta, \phi, t) = \Dr \nabla^2 p(\theta, \phi, t)
\label{eq:ficks_law}
\end{equation}
with the rotational diffusion coefficient $\Dr$.
\begin{figure}
	\centering
	\includegraphics[width = 0.8\columnwidth]{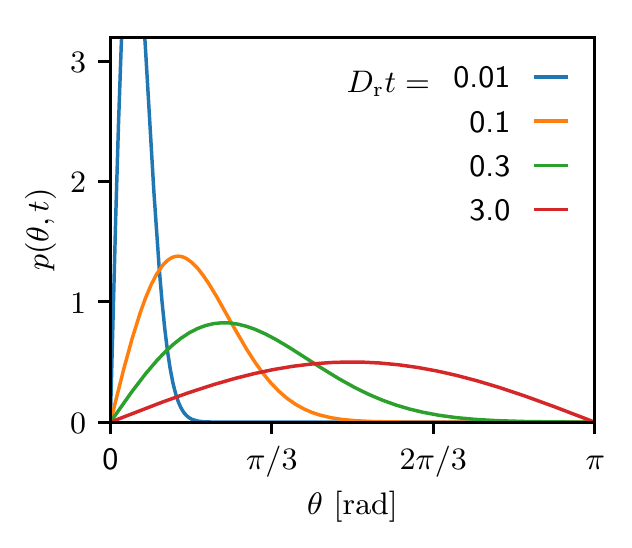}
    \caption{The probability density function of reorientations at various times $t$
    with fixed rotational diffusion coefficient $\Dr$. It shows the time
evolution of the probability density function for the orientation angle
$\theta$.}
	\label{fig:rwos}
\end{figure}

The solution to this equation reads:
\begin{equation}
p(\theta,t) = \sum_{l = 0}^{\infty} \frac{2l+1}{2} e^{-l(l+1)\Dr t} P_l(\cos\theta)\sin\theta,
\label{eq:rwos}
\end{equation}
where $P_l$ is a Legendre polynomial of $l$th order.
Note that in~\cref{eq:rwos} the probability density function $p(\theta, t)$
does not depend on the azimuthal reorientation $\phi$ due to axial
symmetry. This means that $\phi$ can take any value in the range of
$[0,2\pi]$ with equal probability.

\Cref{eq:rwos} states that, as shown in~\cref{fig:rwos}, if the time $t$ for
which a swimmer is allowed to rotate is infinitesimally small, the resultant
reorientation $\theta$ must be infinitesimally small as well. This is because a
swimmer cannot rotate indefinitely fast. The probability density function at an
infinitesimally small time $t$ is thus close to a delta function whose center
is at $\theta = 0$. On the other hand, with increasing rotation time the
reorientation distribution gets broadened.

Obtaining the resulting reorientation distribution, we have hence to weight the
time variable $t$ in~\cref{eq:rwos} by~\cref{eq:geometric_dist} and take an
average over $k$ since each tumble duration is given by the geometrical
distribution function whose termination rate is $\qt$:
\begin{equation}
\begin{aligned}
\mathcal{P}(\theta) &\equiv \left<p(\theta,t)|\mathrm{Pm}(k;q_\tum)\right>_k \\
&= \sum_{l = 0}^{\infty} \frac{2l+1}{2}  P_l(\cos\theta)\sin\theta \\ &\times\sum_{k=1}^{\infty}e^{-\Dr l(l+1)k\delta t}(1-q_\tum)^{k-1}q_\tum.
\end{aligned}
\end{equation}
Note that the time variable $t$ has been replaced by $k\delta t$. Performing
the summation over $k$, we arrive at the \textit{time-independent} weighted
probability density function of reorientations:
%\begin{equation}
%\mathcal{P}(\theta) = \sum_{l = 0}^{\infty} \frac{2l+1}{2}  P_l(\cos\theta)\sin\theta\frac{q_\tum}{e^{\qt l(l+1)\Ro }+q_\tum-1},
%\label{eq:weighted_prob}
%\end{equation}
\begin{equation}
\mathcal{P}(\theta) = \sum_{l = 0}^{\infty} \frac{2l+1}{2}  P_l(\cos\theta)\sin\theta\frac{{\delta t}/{\left<\Tt\right>}}{e^{\Dr l(l+1)\delta t}+{\delta t}/{\left<\Tt\right>}-1}.
\label{eq:weighted_prob}
\end{equation}
\cref{eq:ave_time_resol} has been used to eliminate $\qt$.
%where Ro is a dimensionless number --- we call it a rotation number --- defined as
%\begin{equation}
%\Ro \equiv \Dr\left<\Tt\right>,
%\label{eq:Ro}
%\end{equation}
%and \cref{eq:ave_time_resol} has been used to eliminate $\delta t$.
%The rotation number in~\cref{eq:Ro} captures the fact that ~\cref{eq:weighted_prob} is prescribed neither by $\Dr$
%nor $\left<T_\tum\right>$ but $\Dr \left<T_\tum\right>$. The probability density
%function $\mathcal{P}$ is therefore completely determined by the rotation
%number as well as the Poisson time step $\delta t$ --- although $\mathcal{P}$
%seems to depend on Ro and $\qt$, $\qt$ is subject to Ro through
%$\left<\Tt\right>$, leaving us with $\delta t$ as a free parameter.
%Note that the probability density function $\mathcal{P}$ is therefore completely determined by $\Dr$, $\left<\Tt\right>$, and $\delta t$.
The average value of $\cos\theta$ gives us a better picture of the behavior of $\mathcal{P}$:
\begin{equation}
\begin{aligned}
\left<\cos\theta\right> &= \int_0^{\pi} \cos\theta \mathcal{P}(\theta) \dif \theta \\
&= \sum_{l = 0}^{\infty}\frac{{\delta t}/{\left<\Tt\right>}}{e^{\Dr l(l+1)\delta t}+{\delta t}/{\left<\Tt\right>}-1}\\  &\times \underbrace{\int_0^{\pi} \frac{2l+1}{2}P_l(\cos\theta)\sin\theta\cos\theta \dif \theta}_{\delta_{1l}}\\
& = \frac{{\delta t}/{\left<\Tt\right>}}{e^{2\Dr\delta t}+{\delta t}/{\left<\Tt\right>}-1}.
\end{aligned}
\label{eq:resol_cos}
\end{equation}
%\begin{equation}
%\begin{aligned}
%\left<\cos\theta\right> &= \int_0^{\pi} \cos\theta \mathcal{P}(\theta) \dif \theta \\
%&= \sum_{l = 0}^{\infty}\frac{q_\tum}{e^{\qt l(l+1)\Ro}+q_\tum-1}\\  &\times \underbrace{\int_0^{\pi} \frac{2l+1}{2}P_l(\cos\theta)\sin\theta\cos\theta \dif \theta}_{\delta_{1l}}\\
%& = \frac{q_\tum}{e^{2\qt\Ro}+q_\tum-1}.
%\end{aligned}
%\label{eq:resol_cos}
%\end{equation}
$\left<\cos\theta\right>$ being
non-zero means that the distribution of reorientations is asymmetric.
\Cref{eq:resol_cos} also implies that the Poisson time step $\delta t$ has
influence on the measurement of the reorientation distribution, \textit{i.e.},
the smaller the Poisson time step, the larger $\left<\cos\theta\right>$. The skewness
of the distribution will be discussed in detail in~\cref{sec:results}.

%\begin{figure}
%	\centering
%	\input{plots/ecoli.tex}
%	\caption{A sketch of an \textit{E. coli} bacterium consisting of the body and the flagella.}
%	\label{fig:ecoli}
%\end{figure}
\begin{figure*}[!t]
	\begin{subfigure}[t]{0.33\textwidth}
		\centering
		\includegraphics[scale = 1]{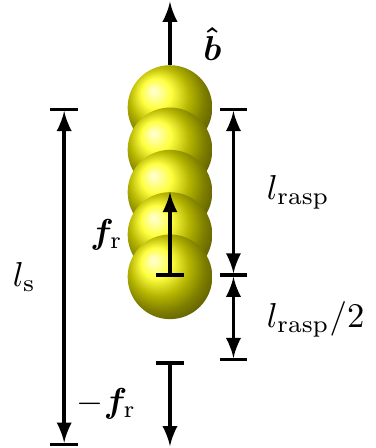}
		\subcaption{}
		\label{fig:run_rasp}
	\end{subfigure}
	\begin{subfigure}[t]{0.33\textwidth}
		\centering
		\includegraphics[scale = 1]{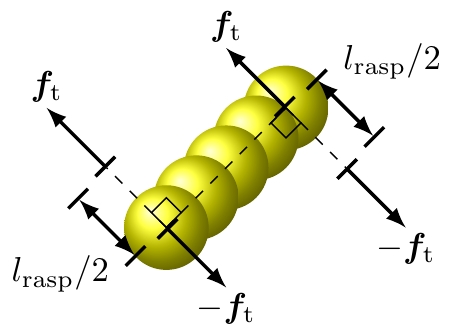}
		\subcaption{}
		\label{fig:tumble_rasp}
	\end{subfigure}
	\begin{subfigure}[t]{0.33\textwidth}
		\centering
		\includegraphics[scale = 1]{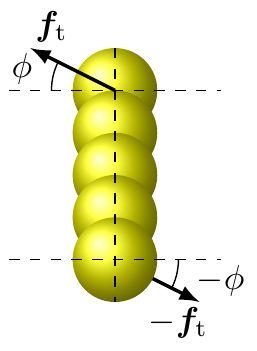}
		\subcaption{}
		\label{fig:azimuthal_def}
	\end{subfigure}
    \caption{A bacterium as a rod-like pusher consisting of 5 point
             particles. (a) representation of the swimmer while it moving along
             $\bm{\hat{b}}$; a force-dipole ($\bm{f}_\mathrm{r}$ and $-\bm{f}_\mathrm{r}$)
             separated by $l_\mathrm{rasp}/2$ is attached to the last point particle. (b) shows
             the tumbling mechanism; two force-dipoles ($\bm{f}_\mathrm{t}$ and
             $-\bm{f}_\mathrm{t}$), each of which is separated by $l_\mathrm{rasp}/2$, are attached to the first and the last point particles, aligned perpendicularly to the swimmer's long axis.
             (c) shows the azimuthal reorientation $\phi$ by which the direction of
             $\bm{f_\mathrm{t}}$ is defined. The dashed line through which $\phi$ is
             specified is arbitrarily chosen but perpendicular to the swimmer's
             long axis. Note that $\bm{\hat{b}}\parallel\bm{f_\mathrm{r}}$ and
             $\bm{f_\mathrm{r}}\perp\bm{f_\mathrm{t}}$ have to be fulfilled.}
    \label{fig:run_and_tumble_rasp}
\end{figure*}
\subsection{Translational diffusion coefficient}\label{subsec:rtm_frm}
Because a RT swimmer's trajectory consists of persistent runs with
sudden changes in direction, the mean-squared displacement (MSD) of
such a swimmer's trajectory is given by~\cite{lovely75a}
\begin{equation}
\left<\Delta r^2\right> = N\left<b^2\right>\left[ \frac{1+\left(2\frac{\left<b\right>^2}{\left<b^2\right>} -1 \right)\left<\cos\theta\right>}{1-\left<\cos\theta\right>}  \right],
\label{eq:msd_rtm_N}
\end{equation}
where $\left< b\right>$ is the average persistent running length,
$\left< b^2 \right>$ the second moment of running lengths $b$, and $N$
the number of persistent runs.

The time it takes for the RT swimmer to complete $N$ number of
persistent runs is approximately~\cite{lovely75a}
$t \approx N\left(\left<T_\tum\right>+ \left<T_\run\right>\right)$.
Note that the subscript $k$ is omitted.  Therefore, the MSD as a
function of time $t$ reads
\begin{equation}
\left<\Delta r^2(t)\right> = \frac{\left<b^2\right>t}{\left<T_\tum\right>+ \left<T_\run\right>}\left[ \frac{1+\left(2\frac{\left<b\right>^2}{\left<b^2\right>} -1 \right)\left<\cos\theta\right>}{1-\left<\cos\theta\right>}  \right].
\end{equation}
Using the definition of the translational diffusion coefficient in
three dimensions, \textit{i.e.}, $\left<\Delta r^2(t)\right> = 6\Dt t$, one can explicitly write
\begin{equation}
\Dt = \frac{1}{6}\frac{\left<b^2\right>}{\left<T_\tum\right>+ \left<T_\run\right>}\left[ \frac{1+\left(2\frac{\left<b\right>^2}{\left<b^2\right>} -1 \right)\left<\cos\theta\right>}{1-\left<\cos\theta\right>}  \right].
\end{equation}
The first two moments, $\left<b\right>$ and $\left<b^2\right>$, can be
calculated using the relations in~\cref{eq:ave_time_resol,eq:geo_std}:
\begin{align}
&\left<b\right> = \left<\Tr\right>\Us =(\left<\Tr\right>+\left<\Tt\right>) \Ua ,\\
&\left<b^2\right> =2(\left<\Tr\right>\Us)^2= 2((\left<\Tr\right>+\left<\Tt\right>)\Ua )^2,
\end{align}
where $\Ua$ is the effective swimming speed of the RT swimmer:
\begin{equation}
\Ua = \frac{\left<\Tr\right>}{\left<\Tr\right>+\left<\Tt\right>}\Us
\end{equation}
with $\Us$ being the swimming speed.

The three expressions above further simplify the determining equation
for the translational
diffusion coefficient:
\begin{equation}
\begin{aligned}
\Dt &= \frac{\Ua ^2}{3}(\left<\Tr\right>+\left<\Tt\right>)\left[ \frac{1}{1-\left<\cos\theta\right>}\right]\\
&=\frac{\Ua ^2}{3}T_\mathrm{c},
\end{aligned}
\label{eq:msd_rtm_final}
\end{equation}
where $\Tc$ is the correlation time, defined as
\begin{equation}
T_\mathrm{c} \equiv \frac{\left<\Tr\right>+\left<\Tt\right>}{1-\left<\cos\theta\right>}\overset{\text{\cref{eq:resol_cos}}}{\approx}2(\left<\Tr\right>+\left<\Tt\right>).
\label{eq:corel_time}
\end{equation}
\Cref{eq:msd_rtm_final} predicts the translational diffusion coefficient from
experimentally accessible quantities. It is worth noting the following
inequality:
\begin{equation}
\Tc \ge \left<\Tr\right>.
\end{equation}
The equality holds if and only if the RT swimmer changes its direction
instantaneously, \textit{i.e.}, $\Tt = 0$, and if the reorientation distribution is symmetric, yielding $\left<\cos\theta\right> = 0$.

\subsection{Rotational diffusion coefficient}
To obtain the rotational diffusion coefficient $\Dr$, we measure the orientational
autocorrelation function \textit{during} tumbles. Since we already know how the reorientation angle $\theta$ evolves in time
from~\cref{eq:rwos}, we can easily calculate the autocorrelation function:
\begin{equation}
\begin{aligned}
\left<\bm{\hat{b}}(t)\cdot\bm{\hat{b}}(0)\right> &= \sum_{l=0}^{\infty}e^{-l(l+1)\Dr t}\\
&\times\int_{0}^{\pi}\frac{2l+1}{2} P_l(\cos\theta)\sin\theta\cos\theta  \dif \theta\\
&=e^{-2\Dr t}.
\end{aligned}
\label{eq:rot_diffu}
\end{equation}
The orientational autocorrelation function shows an exponentially decaying behavior with the exponent being $-2\Dr$.

%\subsection{Clearance in possibly confusing terminologies}
%In molecular dynamics, there are two different diffusion phenomena: the translational diffusivity and the rotational diffusivity. The rotational diffusivity is the same as the inverse correlation time in the sense that both are the measure of directional correlation of an ensemble of trajectories. However, in the run-and-tumble motion (RTM), the rotational diffusion coefficient written in~\cref{eq:ficks_law,eq:rot_diffu} is \textit{not} the reciprocal of the correlation time written in~\cref{eq:corel_time}. We have inevitably introduced another rotational diffusivity in the necessity of describing the time evolution of reorientations. Although, strictly speaking, we have the three different diffusivities, we shall not introduce an additional terminology to refer the rate of the time evolution of reorientations as we can still address the generic rotational diffusion coefficient by the correlation time --- sometimes, it is also called the crossover time. We, therefore, restrict ourselves that the correlation time $\Tc$ always denotes the generic rotational diffusion coefficient and that the rotational diffusion coefficient $\Dr$ denotes the rate of the time evolution of reorientations. For the record, the translational diffusion coefficient $\Dt$ bears the same usual meaning.

\section{Implementation}\label{sec:imple}
In the following we describe a momentum conserving implementation of
the aforementioned run-and-tumble statistics for a hybrid MD/LB
simulation within the software package
\textsf{ESPResSo}~\cite{weik19a,arnold13a}. The lattice-Boltzmann
method serves as a hydrodynamics solver~\cite{succi01a,krueger17a},
whereas the molecular dynamics method solves Newton's equations of
swimmer's motion. These two simulation methods are coupled via a
frictional coupling scheme described in
Ref. \citet{ahlrichs99a}. Including hydrodynamic interactions makes
our simulation scheme not only versatile as it allows us to study
multiple RT swimmers without sacrificing the swimmers'
hydrodynamic interactions, but also satisfy the momentum conservation
law (see below). Note that one can also exclude the hydrodynamical
interactions by simply not using the LB, since
our RT algorithm does not rely on the hydrodynamical interactions. The model
can, therefore, be implemented also in other numerical schemes,
\textit{e.g.}, like Langevin or Brownian MD, or Monte Carlo.

\subsection{Swimmer configuration}
\begin{figure}
	\begin{subfigure}[t]{0.45\columnwidth}
		\centering
		\includegraphics[width = \columnwidth]{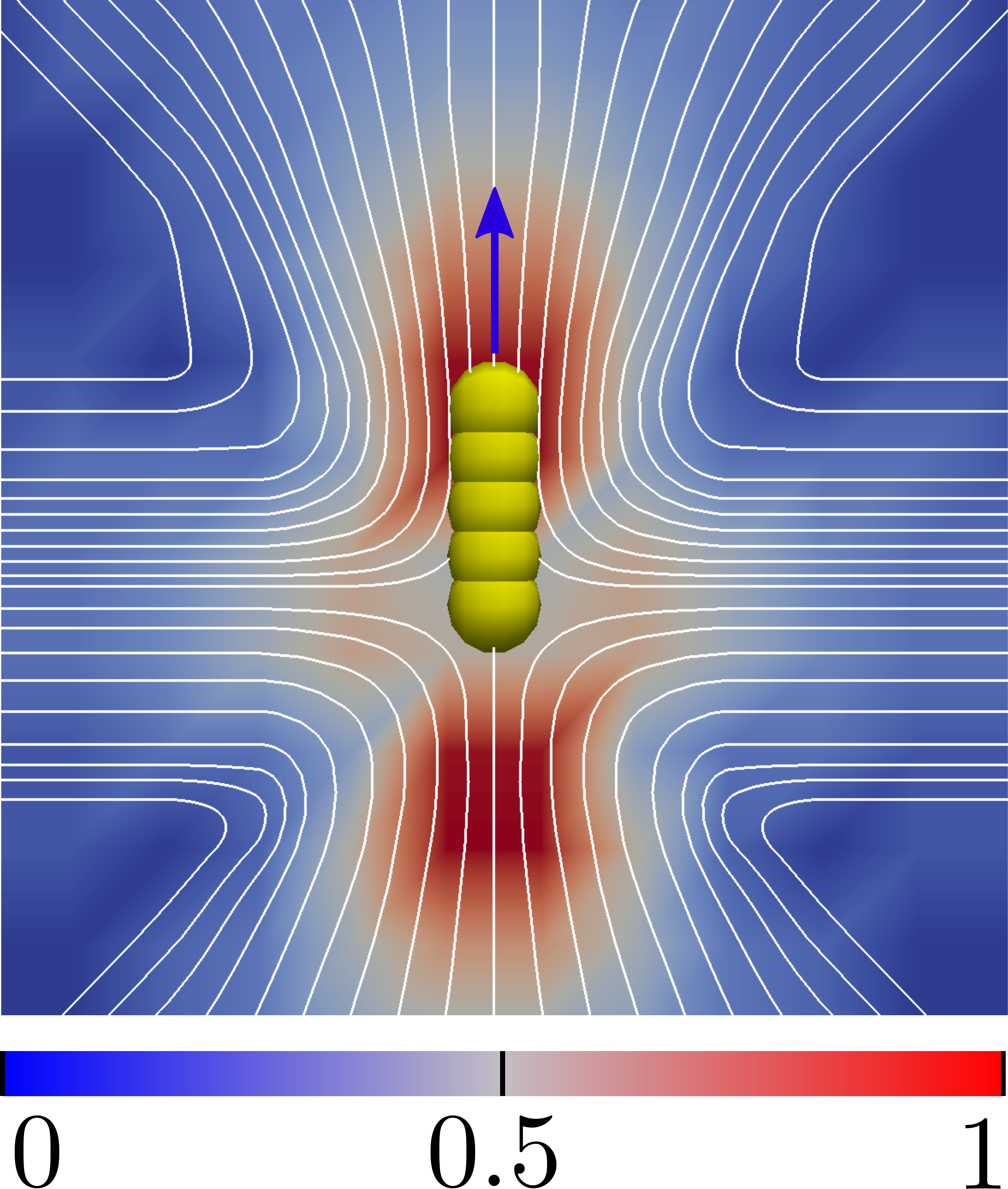}
		\subcaption{}
		\label{fig:run_rasp_sim}
	\end{subfigure}
	\begin{subfigure}[t]{0.45\columnwidth}
		\centering
		\includegraphics[width = \columnwidth]{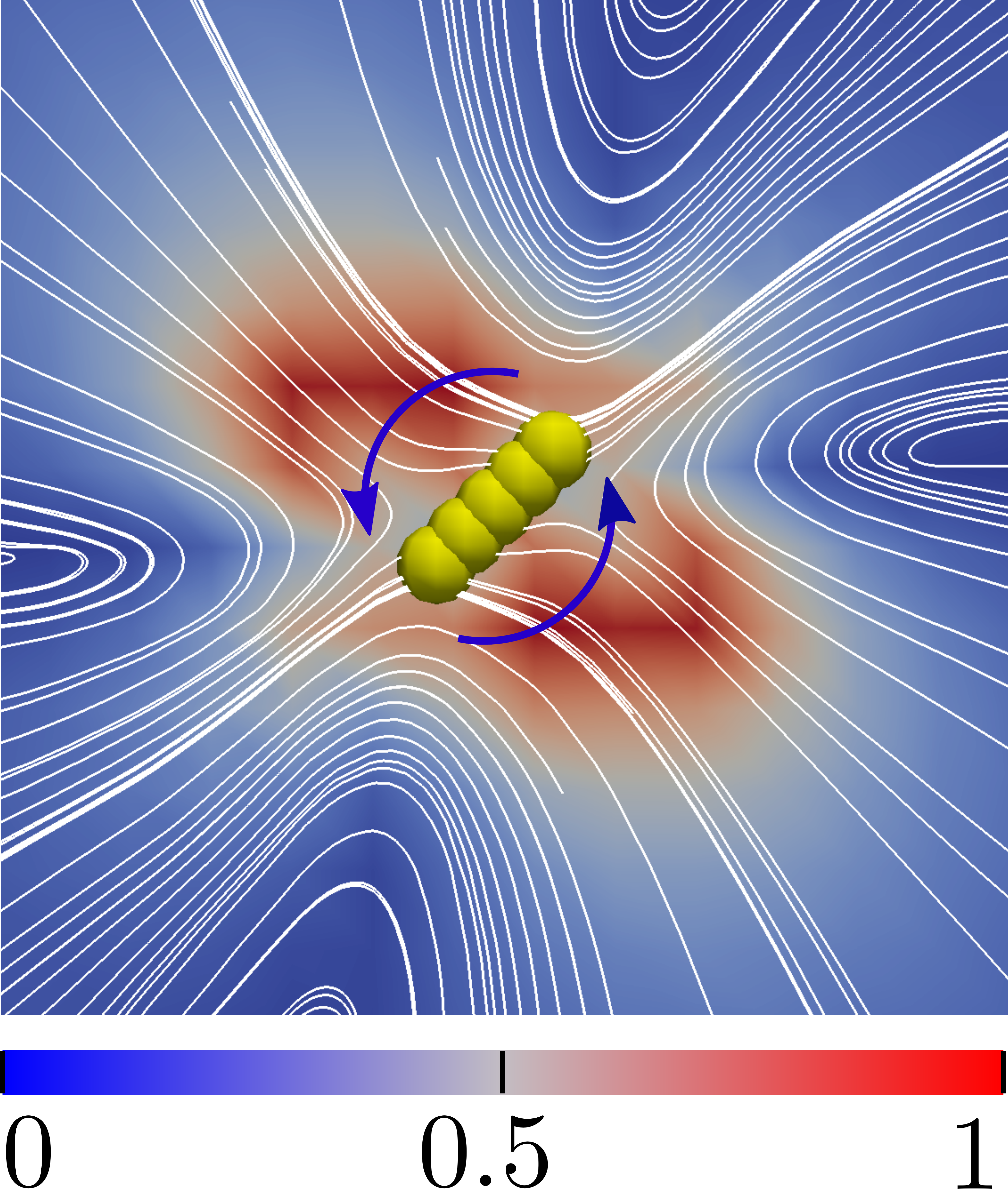}
		\subcaption{}
		\label{fig:tumble_rasp_sim}
	\end{subfigure}
    \caption{The induced flow field of the swimmer in the
laboratory frame during (a) running, and (b) tumbling, respectively. The color
map represents the normalized flow speed, and the white lines the streamlines
which massless tracers would follow.}
	\label{fig:ff_run_and_tumble}
\end{figure}
\begin{figure}
	\centering
	\includegraphics[width = 0.7\columnwidth]{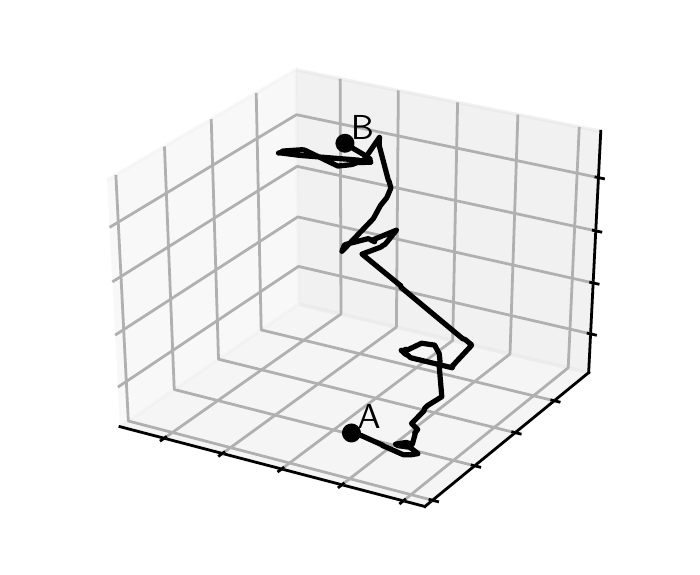}
    \caption{The trajectory of the run-and-tumble swimmer in the simulation.
The point A is the starting position, and the point B is the position at time
$10^{7}\tau$.}
	\label{fig:rtm_traj}
\end{figure}
\begin{figure}
	\centering
	\includegraphics[width = \columnwidth]{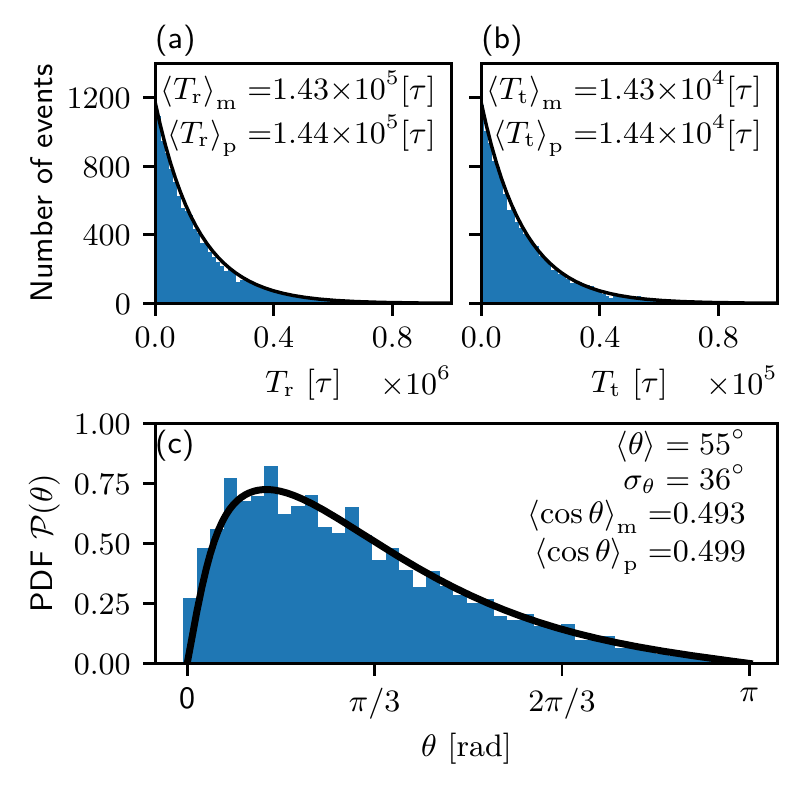}
    \caption{The durations of runs (a) and tumbles (b), and the distribution of
reorientations (c). $\left<\square\right>_\mathrm{m}$ and $\left<\square\right>_\mathrm{p}$ indicate an average value that we \textit{measure} and an average value that we \textit{predict} from the parameters that we prescribed, respectively. The black lines represent the corresponding predicted curves.}
	\label{fig:hist_rtm}
\end{figure}
\begin{figure*}
	\centering
	\includegraphics[width = \textwidth]{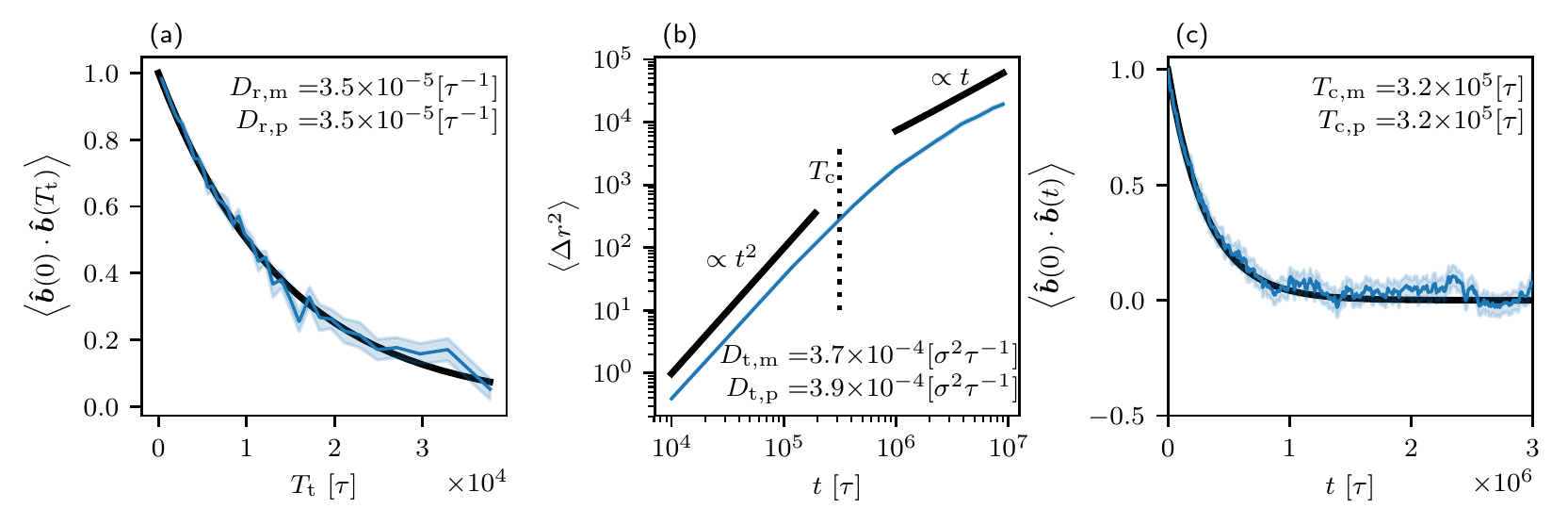}
        \caption{(a): The orientational autocorrelation as a function
          of tumble duration $\Tt$. The black curve is the fit from
          which $D_\mathrm{r,m}$ is taken. (b): The mean-squared
          displacement (MSD) of the swimmer ensemble 
          trajectory. The black lines denote the fits for the
          ballistic ($\propto t^2$) and diffusive ($\propto t$)
          regimes. The translational diffusion coefficient $D_\mathrm{t,m}$ is
          obtained by a linear fit to the diffusive regime. The
          vertical dashed line indicates the correlation time
          $\Tc$. (c): The directional correlation as a function of
          time $t$, measuring the correlation time $\Tc$. The black curve is the fit from which to obtain $T_\mathrm{c,m}$. Throughout all plots, the shadowed zones covers the corresponding standard errors.}
	\label{fig:diffu}
\end{figure*}

If we would couple the particle only at one point to the LB fluid we
would not be able to prescribe a torque on it. Therefore we use a
raspberry
approach~\cite{lobaskin04a, fischer15a,degraaf15b,degraaf16a}. For our
model we construct the swimmer as a rigid body consisting of 5 point
particles. The total length of the swimmer is denoted by
$l_\mathrm{rasp}$ as shown in~\cref{fig:run_and_tumble_rasp}. The
effective diameter of each particle is set to be $l_\mathrm{rasp}/2$,
which is realized through a Weeks-Chandler-Andersen interaction
potential~\cite{weeks71a}. The raspberry method is particularly
useful for modeling arbitrarily shaped objects due to its great
simplicity and versatility.

The model bacterium has now acquired some internal structure and thus two rotational
degrees of freedom: rotating and precessing. Note that the swimmer still cannot
experience a torque that spins it with respect to the swimmer's long axis,
which corresponds to the change in azimuthal angle $\phi$. Therefore, the
change in $\phi$ happens instantaneously (see~\cref{fig:azimuthal_def}).

A force is applied onto the swimmer during the running phase in order
to model the swimming mechanism. By applying the corresponding
counter-force onto the fluid at a distance of $l_\mathrm{rasp}/2$ the
total linear momentum is conserved. These two forces thus form a
force-dipole, turning the particle into a so-called pusher
swimmer~\cite{degraaf16a,degraaf16b}(see~\cref{fig:run_rasp_sim}). Keep
in mind that the distance between the force and counter-force, or the
dipole distance, is not fixed by requiring momentum conservation
alone. It is a reasonable choice to fix it somewhere within the length
of the bacterias flagella, and it should not be too small since then
the flow field generated by the counter-force starts affecting the
dynamics of the swimmer~\cite{degraaf16a}. Therefore, we have chosen a
dipole distance of $l_\mathrm{rasp}/2$.

Note that a force-dipole exhibits two singularities, and a swimmer
experiences a repulsive hydrodynamic force by approaching another
swimmer's "tail", where the counter-force is being applied to the
fluid~\cite{zoettl16a,winkler16a,spagnolie12a}. Therefore, the
effective size of the swimmer can be approximated by twice the length
of the raspberry particle, that is,
$l_\mathrm{s} \approx 2l_\mathrm{rasp}$\footnote{In fact, the size of
  swimmer is irrelevant for a single swimmer if there are no other
  object with which the swimmer interacts. However, this notion is
  still useful because a (swimming) P\'eclet number, describing the
  persistency of swimmer's directional motion, requires a length scale
  (see~\cref{subsec:pram}).}.

During tumbling we attach two oppositely pointing force-dipoles at the
two terminating particles of the swimmer, aligned perpendicularly to the
swimmer's long axis (see~\cref{fig:tumble_rasp}). This always
guarantees angular momentum conservation when the swimmer is
tumbling, regardless of the dipole distances. Each force-dipole is,
due to the same reason mentioned above, separated by
$l_\mathrm{rasp}/2$ as well. Note that the direction of the
force-dipoles on the azimuthal plain are defined by an arbitrarily
chosen axis that is perpendicular to the swimmer's long axis
(see~\cref{fig:azimuthal_def}).

The corresponding flow fields when the swimmer is running and tumbling are
shown in~\cref{fig:ff_run_and_tumble}. Note that the force-dipole scheme is one
of the simplest representations of bacteria inducing flow fields that satisfy
the momentum conservation law. Thus, we want to stress that this model
only reproduces the correct far
field~\cite{winkler16a,zoettl16a,spagnolie12a}, but due to its
asymmetric shape it can also be influenced hydrodynamically via flows
like any similarly shaped bacterium.

\subsection{Simulation parameters of the run-and-tumble algorithm}\label{subsec:pram}
We need to set several parameters beforehand: the swimmer's length
$l_\mathrm{s}$, the average durations of runs $\left<\Tr\right>$ and
tumbles $\left<\Tt\right>$, the Poisson time step $\delta t$, the
swimming speed $\Us$, and the rotational diffusion coefficient
$\Dr$. Here, we aim to reproduce the dynamics of \textit{E. coli} as
closely as possible to the experimental data found in
Refs.~\citet{berg72a,berg93a,saragosti12a}.

The parameters that are related to the swimmer's running motion are
subject to a (swimming) P\'eclet number, which is defined
as~\cite{zoettl16a,clement16a}:
\begin{equation}
\Pec = \frac{\Ua T_\mathrm{c}}{l_\mathrm{s}} \approx \frac{2 \Us\left<\Tr\right>}{l_\mathrm{s}}.
\label{eq:Pec}
\end{equation}
For the sake of comparability against the experiments, we set Pe as
4.8, which is obtained from the corresponding experimental
data~\cite{berg72a,berg93a,saragosti12a} with
$l_\mathrm{s}=10\mathrm{\mu m}$~\cite{clement16a}. Note that, however,
the P\'eclet number can be different from one experiment to another
since the quantities in~\cref{eq:Pec} can show large variances.

The swimmer's tumbling motion, on the other hand, is governed by
the rotational diffusion coefficient $\Dr$ and the average tumble duration $\left<\Tt\right>$. We set $\Dr = 5s^{-1} = 3.5\times 10^{-5}\tau^{-1}$ and $\left<\Tt\right> = 0.1s = 1.44\times10^4\tau$ with $\tau$ being the LB-MD time step. These values are again taken from the
experiments~\cite{berg72a,berg93a,saragosti12a,clement16a}.
%When it comes to the swimmer's tumbling motion, the parameters are
%subject to a rotation number defined in~\cref{eq:Ro}.  Ro is
%prescribed as 0.5, which is again taken from the
%experiments~\cite{berg72a,berg93a,saragosti12a,clement16a}.

The only free parameter left is the Poisson time step which defines
the accuracy of the run and tumble durations. Therefore, the Poisson
time step should be small compared to $\left<\Tt\right>$. However, if the Poisson time
step gets smaller the simulation becomes computationally more
expensive. We thus found a reasonable balance between accuracy and
computational speed at $\delta t = 100\tau$.

Once these parameters are determined, we iteratively apply the following
scheme:
\begin{enumerate}
\item{Draw a random number for a running duration $T_\run$ from the
    geometric distribution~\cref{eq:geometric_dist} whose termination
    rate is $q_\run$.}\label{step:1}
	\item{Let the swimmer run with the swimming speed $\Us $ for $T_\run$.}
	\item{Draw a random number for a tumbling duration $T_\tum$
            from the geometric distribution~\cref{eq:geometric_dist}
            whose termination rate is $q_\tum$.}
	\item{With the randomly drawn tumbling duration $T_\tum$, draw
            a random number for a reorientation $\theta$ from the
            probability density function~\cref{eq:rwos}.}
        \item{Draw a random number for an azimuthal reorientation
            $\phi$ from a uniform distribution that ranges $[0,2\pi]$,
            and make the swimmer azimuthally "spin" by $\phi$.}
	\item{Assign the angular speed $\Omega = \theta/T_\tum$ to the
            swimmer, and let it rotate for $T_\tum$.}
	\item{Go back to the step 1.}
\end{enumerate}

\section{Results}\label{sec:results}

We demonstrate the validity of our run-and-tumble swimmer model by
analyzing the following observables: the durations of runs and
tumbles, the distribution of reorientations, and the translational and
rotational diffusion coefficients.

We place a swimmer in a periodic cubic box whose side length is
$40\sigma$ with $\sigma$ being the diameter of a particle
($\sigma = l_\mathrm{rasp}/2$). The box length is chosen to be large
enough such that any artifacts due to periodic boundary conditions are
negligible. Initially the lattice Boltzmann fluid is set up in
equilibrium. We ran 20 independent simulations for $10^{8}\tau$. To make a reasonable ensemble set,
we cut each simulation into 10 blocks. We then construct the ensemble
with 200 independent data sets.  The chopping does not compromise the
quality of the data since $10^{7}\tau$ is long enough for the system
to be uncorrelated with its initial state, that is,
$\Tc \ll 10^{7}\tau$. A sample for a RT trajectory is shown
in~\cref{fig:rtm_traj}.

We first analyze the distributions of the RT motion, and our results
agree with the experimental data~\cite{berg72a,berg93a}. As stated
in~\cref{subsec:dist}, the duration distributions of runs and tumbles
in~\cref{fig:hist_rtm}a and~\cref{fig:hist_rtm}b follow a Poisson
statistics. The measured distribution of reorientations
in~\cref{fig:hist_rtm}c matches with the analytically formulated
probability function (see~\cref{eq:weighted_prob}).

The important remark here is the skewed distribution of
reorientations, which was first discovered by~\citeauthor{berg72a} in
1972~\cite{berg72a}. This is because the most probable duration of
tumbles is very close to 0 as shown in~\cref{fig:hist_rtm}b. This
leads to the most probable reorientation to be very small as
well. Consequently, the resultant distribution of reorientations is
skewed to a smaller angle.

Note that, as mentioned in~\cref{sec:theory}, the Poisson time step affects the measure of $\left<\cos\theta\right>$, and our choice of the Poisson time step is
100$\tau$. With this value, we
analytically predict $\left<\cos\theta\right>=0.499$ and measured
$\left<\cos\theta\right>=0.493$ (see~\cref{fig:hist_rtm}c). In the limit of an infinitely small Poisson time step, the expected value is
$\lim\limits_{\delta t \to 0}\left<\cos\theta\right>=0.5$.

What is of equal importance as the distributions are the rotational
and translational diffusion coefficients. In~\cref{fig:diffu}a, we
display our data for the rotational diffusion coefficient $\Dr$ that
turns out to be very close to the assigned value of $\Dr$. This is one
of the indications that our simulation algorithm works as
intended. When it comes to the translational diffusion coefficient
$\Dt$, however, we have \textit{not} prescribed it. $\Dt$ results
solely from the ensemble of our swimmer's trajectories. We have
reproduced \textit{E. coli}'s run-and-tumble motion within 5\% of the
relative error judging by the translation diffusion coefficient
(see~\cref{fig:diffu}b). In addition, one can clearly see the
characteristic behavior of an MSD: a ballistic regime for short times
and a transition to a diffusive regime for longer times. The
transition happens around the correlation time $\Tc$, which is also
precisely captured by our data, as shown in~\cref{fig:diffu}c.

It is worth mentioning the discrepancy between the experimental results obtained by
\citeauthor{berg93a} and ours.  He measured an average angle of
$\left<\cos\theta\right>\sim 0.33$~\cite{berg93a}, which is smaller than what
we measured. We identified three possible reasons for this difference, namely
the introduction of a threshold angle for reorientations in the experiment,
the uncertainty in the rotational diffusion constant $\Dr$ and finally the
finite frame rate of the recording device in the experiment of \citeauthor{berg93a}.

\section{Conclusion and outlook}

In summary, we have implemented an algorithm for the RT
motion of a self-propelled particle coupled to a lattice-Boltzmann
fluid. Furthermore we have developed an expression for the
time-independent distribution of reorientations describing the
RT motion of bacteria. With the help of this expression
we analyzed an ensemble of RT trajectories, obtained via
our LB-MD simulations of a single RT swimmer, which are
prescribed by a P\'eclet number (Pe), a rotational diffusion coefficient ($\Dr$), an average tumble duration ($\left<\Tt\right>$), and a
Poisson time step ($\delta t$).

Our RT swimmer model reproduces the real
\textit{E. coli}'s RT motion with excellent accuracy. Our analysis of the
mean-squared displacement further demonstrates that our model provides
the correct translational and rotational diffusion constants of an
\textit{E. coli} bacterium. Another advantage of our model over a
standard Langevin implementation is that, apart from incorporating
hydrodynamics, we have an explicit reorientation process of
the RT motion, which reproduce nearly identical
trajectories to those of \textit{E. coli}.

In future investigations we will use this implementation to study the
collective behavior of \textit{E. coli} in various environments and
under external flow conditions.

\begin{acknowledgments}
  We thank the Deutsche Forschungsgemeinschaft (DFG) for funding our
  research through the SPP 1726 ``Microswimmers: from single particle
  motion to collective behavior'' (HO1108/24-2) and through the CRC
  1313 ``Grenzfl\"achenbeeinflusste Mehrfeldprozesse in por\"osen
  Medien - Str\"omung, Transport und Deformation'', Research Project
  C.1, and we would like to acknowledge inspiring discussions with
  J. de Graaf, M. Kuron, G. Rempfer, and C. Lohrmann.
\end{acknowledgments}

\section*{References}
\bibliography{bibtex/icp}

%\begin{figure}[t]
%	\centering
%	\inputpgf{plots/}{analytic_run_and_tumble.pgf}
%	\caption{(a) and (b) represent the probability mass functions of tumbles and runs with the termination rates $q_\tum = 10^{-3}$ and $q_\run = 10^{-4}$, respectively.  (c) shows the probability density function of reorientations. The colored lines, starting with the blue one, denote the time evolution of this function, provided that $\Dr $ is fixed. The black line depicts the weighted probability density function $\mathcal{P}(\theta)$ (see~\cref{eq:weighted_prob}), which corresponds to~\cref{fig:rtm_duration_berg} (right).}
%	\label{fig:analytic_rtm}
%\end{figure}

%\begin{figure}
%	\centering
%	\input{plots/frm.tex}
%	\caption{A schematic trajectory of a run-and-tumble bacterium with 5 tumble events.}
%	\label{fig:frm}
%\end{figure}

\end{document}